\documentstyle{mn}
\input{epsf}

\newif\ifAMStwofonts

\newcommand{\qniec}{\end{document}}
\newcommand{\etal}{{et al.}~}
\newcommand{\de}{\delta}
\newcommand{\te}{\theta}

\newcommand{\Sig}{\Sigma}
\newcommand{\lam}{\lambda}
\newcommand{\p}{\partial}
\newcommand{\f}{\frac}
\newcommand{\s}{\sigma}
\newcommand{\bfx}{\bmath{x}}

\newcommand{\bfv}{\bmath{v}}

\newcommand{\calO}{{\mathcal O}}

\newcommand{\calK}{{\mathcal K}}
\newcommand{\calN}{{\mathcal N}}
\newcommand{\calS}{{\mathcal S}}

\newcommand{\bc}{\begin{center}}
\newcommand{\be}{\begin{equation}}
\newcommand{\ee}{\end{equation}}
\newcommand{\ec}{\end{center}}
\newcommand{\lan}{\langle}
\newcommand{\ran}{\rangle}

\newcommand{\hmpc}{\, h^{-1}\rmn{Mpc}}
\newcommand{\kms}{{$\rmn{km}\; \rmn{s}^{-1}$}}

\ifoldfss
  \newcommand{\rmn}[1] {{\rm #1}}

  \ifCUPmtlplainloaded \else
    \NewTextAlphabet{textbfit} {cmbxti10} {}
    \NewTextAlphabet{textbfss} {cmssbx10} {}
    \NewMathAlphabet{mathbfit} {cmbxti10} {} 
    \NewMathAlphabet{mathbfss} {cmssbx10} {} 
  \fi
  \ifAMStwofonts
    \ifCUPmtlplainloaded \else
      \NewSymbolFont{upmath} {eurm10}
      \NewSymbolFont{AMSa} {msam10}
      \NewMathSymbol{\upi}     {0}{upmath}{19}
      \NewMathSymbol{\umu}     {0}{upmath}{16}
      \NewMathSymbol{\upartial}{0}{upmath}{40}
      \NewMathSymbol{\leqslant}{3}{AMSa}{36}
      \NewMathSymbol{\geqslant}{3}{AMSa}{3E}

       \let\le=\leqslant
       
    \fi
  \fi
\fi 

\ifnfssone
  \newmathalphabet{\mathit}
  \addtoversion{normal}{\mathit}{cmr}{m}{it}
  \addtoversion{bold}{\mathit}{cmr}{bx}{it}
  \newcommand{\rmn}[1] {\mathrm{#1}}

  \newmathalphabet{\mathbfit} 
  \addtoversion{normal}{\mathbfit}{cmr}{bx}{it}
  \addtoversion{bold}{\mathbfit}{cmr}{bx}{it}
  \newmathalphabet{\mathbfss} 
  \addtoversion{normal}{\mathbfss}{cmss}{bx}{n}
  \addtoversion{bold}{\mathbfss}{cmss}{bx}{n}
  \ifAMStwofonts
    \ifCUPmtlplainloaded \else
      \UseAMStwoboldmath
      \makeatletter
      \new@mathgroup\upmath@group
      \define@mathgroup\mv@normal\upmath@group{eur}{m}{n}
      \define@mathgroup\mv@bold\upmath@group{eur}{b}{n}
      \edef\UPM{\hexnumber\upmath@group}
      \new@mathgroup\amsa@group
      \define@mathgroup\mv@normal\amsa@group{msa}{m}{n}
      \define@mathgroup\mv@bold\amsa@group{msa}{m}{n}
      \edef\AMSa{\hexnumber\amsa@group}
      \makeatother
      \mathchardef\upi="0\UPM19
      \mathchardef\umu="0\UPM16
      \mathchardef\upartial="0\UPM40
      \mathchardef\leqslant="3\AMSa36
      \mathchardef\geqslant="3\AMSa3E

       \let\le=\leqslant

    \fi
  \fi
\fi 

\ifnfsstwo
  \newcommand{\rmn}[1] {\mathrm{#1}}

  \DeclareMathAlphabet{\mathbfit}{OT1}{cmr}{bx}{it}
  \SetMathAlphabet\mathbfit{bold}{OT1}{cmr}{bx}{it}
  \DeclareMathAlphabet{\mathbfss}{OT1}{cmss}{bx}{n}
  \SetMathAlphabet\mathbfss{bold}{OT1}{cmss}{bx}{n}
  \ifAMStwofonts
    \ifCUPmtlplainloaded \else
      \DeclareSymbolFont{UPM}{U}{eur}{m}{n}
      \SetSymbolFont{UPM}{bold}{U}{eur}{b}{n}
      \DeclareSymbolFont{AMSa}{U}{msa}{m}{n}
      \DeclareMathSymbol{\upi}{0}{UPM}{"19}
      \DeclareMathSymbol{\umu}{0}{UPM}{"16}
      \DeclareMathSymbol{\upartial}{0}{UPM}{"40}
      \DeclareMathSymbol{\leqslant}{3}{AMSa}{"36}
      \DeclareMathSymbol{\geqslant}{3}{AMSa}{"3E}

       \let\le=\leqslant

    \fi
  \fi
\fi 

\ifCUPmtlplainloaded \else
  \ifAMStwofonts \else 
    \def\upi{\pi}
    \def\umu{\mu}
    \def\upartial{\partial}
  \fi
\fi

\title[Cosmological peculiar velocity from the density field]
{Recovery of the cosmological peculiar velocity from 
the density field in the weakly non-linear regime}
\author[Chodorowski, {\L}okas, Pollo and Nusser]
       {M.~J.~Chodorowski,$^1$  
	E.~L.~{\L}okas,$^1$    
	A.~Pollo$\,^1$ 
	and A.~Nusser$\,^2$\\	
 	$^1$Copernicus Astronomical Center, Bartycka 18, 
	00--716 Warsaw, Poland\\
	$^2$Max-Planck-Institut f\"ur Astrophysik, 
	Karl-Schwarzschild-Str.\ 1, 85740 Garching, Germany 
	}

\begin{document}

\maketitle

\begin{abstract}
Using third-order perturbation theory, we derive a relation between
the mean divergence of the peculiar velocity given density and the
density itself. Our calculations assume Gaussian initial conditions
and are valid for Gaussian filtering of the evolved density and
velocity fields. The mean velocity divergence turns out to be a
third-order polynomial in the density contrast. We test the power
spectrum dependence of the coefficients of the polynomial for
scale-free and standard CDM spectra and find it rather weak. Over
scales larger than about $5 \hmpc$, the scatter in the relation is
small compared to that introduced by random errors in the observed
density and velocity fields. The relation can be useful for recovering
the peculiar velocity from the associated density field, and also for
non-linear analyses of the anisotropies of structure in redshift
surveys.

\end{abstract}


\section{Introduction}
\label{sec-intro}
In the gravitational instability paradigm for the formation of
structure in the Universe, the peculiar motions (deviations from the
Hubble flow) of galaxies are tightly related to the mass distribution
in the Universe. Quantitative relations between the peculiar velocity
field, $\bfv$, and the mass density contrast field, $\de = \rho /
\rho_b -1$, where $\rho_b$ is the background density, can be obtained
from the  equations of motion of a collisionless self-gravitating
system.  In the linear regime, the relation between the density
and the velocity fields is \be \de(\bfx) = - f(\Omega)^{-1} \nabla
\cdot \bfv(\bfx) \,,
\label{e1}
\ee 
where $f(\Omega) \simeq \Omega^{0.6}$ (e.g.\ Peebles 1980) and we
express distances in units of \kms. Given an assumed value for
$\Omega$, the above relation can be used to reconstruct the mass
density field from the large-scale velocity field.  The comparison of
the reconstructed mass field with the observed large-scale galaxy
density field serves as a method for estimating $\Omega$ (Dekel \etal
1993) and as a test of the gravitational instability hypothesis (but
see Babul \etal 1994).  The linear theory relation is applicable only
when the density fluctuations are small compared to unity. However,
the observed density fluctuations from current redshift surveys
(e.g. Fisher \etal 1994) and from the {\sc potent} reconstruction of 
density fields, slightly exceed the regime of applicability of linear
theory. For example, the density contrast in regions like the Great
attractor is about unity even when smoothed over scales of the order
of $1000$ \kms\ (Dekel \etal 1993).  Future redshift surveys and
catalogs of peculiar velocities will provide reliable estimates of
density and velocity fields on scales where nonlinear effects are
certainly not negligible and need to be incorporated in analyzing the
data.  Various nonlinear relations between the velocity and the
associated density field have been developed. One approach is to
assume phenomenological parametric forms of these relations, which are
to be calibrated with N-body simulations (Dekel 1994, Ganon \etal 1998). 
Another complementary approach is to analytically derive 
these relations based on various approximations to nonlinear dynamics. 
For the purpose of deriving the density from the observed velocities, the
{\sc potent} algorithm for example uses a nonlinear approximation based on
the Zel'dovich approximation (Nusser \etal 1991). The inverse problem,
in which one wishes to recover the velocity from a given density field
is yet unsolved in the Zel'dovich approximation as it involves a set
of nonlinear differential equations, which do not have analytic
solutions.  This paper aims at deriving a nonlinear relation which can
be used to recover the velocity field from the density field. This is
relevant to comparisons of observed velocities of galaxies, using for
example the Tully-Fisher relation, with predicted velocity fields from
the density field estimated from redshift surveys (Strauss \& Davis
1988, Yahil 1988, Kaiser \etal 1991, Hudson 1994, Davis, Nusser \&
Willick 1996, Willick \& Strauss 1998). Moreover, such a relation can
be useful for estimating $\Omega$ from the distortions of clustering
in redshift space (Kaiser 1987, Fisher \& Nusser 1996, Taylor \&
Hamilton 1996, Hamilton 1997) when measured, for example, from the
anisotropies of the correlation functions or, equivalently, the power
spectra. This is particularly important for measurements of redshift
distortions in the Sloan Digital Sky Survey (Gunn \& Knapp 1993) and
the Anglo-Australian 2dF galaxy survey. Nonlinear analysis of these
data might also aim at breaking the degeneracy between $\Omega$ and
bias (Chodorowski \& {\L}okas 1997; hereafter Paper~I, Bernardeau,
Chodorowski \& {\L}okas, 1998).

One way to go in order to derive velocity from density is to assume
that the divergence of the velocity field at any point in space can be
approximated by an expansion in terms of the density contrast at that
point. The coefficients of this expansion can then be determined using
either N-body simulations (Nusser \etal 1991, Mancinelli \etal 1994)
or perturbation theory (Bernardeau 1992) under the assumption of
Gaussian initial conditions. The advantage of this approach is that
it provides the velocity divergence directly in terms of the
density. For irrotational flows, the velocity field can readily be
recovered from its divergence given some boundary conditions at large
distances. Indeed, if we define\footnote{Note a slight difference
from the commonly used definition, e.g.\ Bernardeau~(1994)} \be
\te(\bfx)\equiv - f(\Omega)^{-1} \nabla\cdot\bfv(\bfx) ,
\label{e2}
\ee
then the velocity field is simply
\be
\bfv(\bfx) = \f{f(\Omega)}{4\pi} \int {\rm d}^3 x' \te(\bfx')
\f{\bfx'- \bfx}{\vert \bfx'- \bfx \vert^3} \, .
\label{e6}
\ee 
One caveat to this approach is that the velocity divergence is not
determined uniquely by the density field (Chodorowski 1997, Mancinelli
\& Yahil 1995, Catelan \etal 1995) and the scatter around the derived
relation is expected to propagate into errors in the velocity field.
However, in the weakly nonlinear regime, the values of $\de(\bfx)$ and
$\te(\bfx)$ are still strongly correlated (Bernardeau~1992, Paper~I)
and the merits of this approach completely overwhelm this caveat given
that the problem is not deterministically solved even in the simple
Zel'dovich approximation. 

In this paper we rigorously compute the mean $\te(\bfx)$ given
$\de(\bfx)$, i.e., $\lan \te \ran|_{\de}$, up to third order in
perturbation theory, assuming Gaussian initial conditions.  As we
shall see later, the derivation is greatly simplified when following
Paper~I in which the mean of $\de(\bfx)$ given $\te(\bfx)$ was
calculated. In Paper I we explicitly calculated the coefficients,
$a_i$, appearing in the expansion

\be 
\Delta(\te) \equiv \lan \de \ran|_{\te} = 
a_1 \te + a_2 (\te^2 - \s_\te^2) + a_3 \te^3 \, ,
\label{e3}
\ee
where $\s_\te^2$ is the variance of the velocity divergence field.
The numerical values of $a_i$ are found to be in good agreement with
the results of N-body simulations by Chodorowski \etal (1998) and by
Ganon \etal (1998). The relation~(\ref{e3}) allows one to reconstruct
a density field from the corresponding velocity field.

But why can we not invert the relation~(\ref{e3}) to obtain the
$\Theta(\de) \equiv \lan \te \ran|_{\de}$ to the relevant order in
perturbation theory? The reason is simply that the scatter of $\de$
around the mean value given by~(\ref{e3}) introduces, in general, a
bias in the estimate of $\Theta(\de)$ obtained by straightforward
inversion.

The paper is organized as follows: in Section~\ref{sec-form} we
describe how the mean of the velocity divergence can be computed
from the density using the results of Paper I. In
Section~\ref{sec-num} we calculate the numerical values of the
coefficients entering the formula for $\Theta(\de)$.
Section~\ref{sec-scatter} is devoted to the computation of the scatter
in the $\te$--$\de$ relation. Summary is given in
Section~\ref{sec-summ}. 

\section{The formalism}
\label{sec-form}
We now outline the derivation of the relation between the divergence
of the velocity field and the density contrast.  Let us express the
density contrast as a sum of terms $\delta_i$, each corresponding to
the $i^{th}$ order in perturbation theory,

\be 
\de = \de_{1} + \de_{2} + \de_{3} + \ldots \,,
\label{e9}
\ee
and, similarly, for the velocity divergence

\be \te = \te_{1} + \te_{2} + \te_{3} + \ldots \,.
\label{e10}
\ee

In general, the $i^{th}$ order solution is of the order of $(\de_1)^i$
(Fry 1984, Goroff \etal 1986). We assume here that $\te$
and $\de$ are well approximated by truncating the above expansion at
the third order. The linear theory solution mentioned in Section~1 is
simply the perturbative expansion truncated at the lowest, i.e.~first
order term, $\de_1$ and $\te_1$. (For explicit forms of higher-order
solutions and other details see Paper~I.) 
The coefficients, $a_i$, appearing in the expansion of the
function $\Delta$, equation~(\ref{e3}), are combinations of the
joint moments of the density and velocity divergence fields. According
to Paper~I
\begin{eqnarray}
a_1
&=& 1 + \left[ \Sigma_2 + \f{(S_{3\de} - S_{3\te})\, S_{3\te} }{3} -
\f{\Sigma_4}{2} \right] \s^2 \,,
\label{e11} \\
a_2
&=& \f{S_{3\de} - S_{3\te}}{6} \,,
\label{e12} \\
a_3
&=& \f{\Sigma_4 - (S_{3\de} - S_{3\te})\, S_{3\te} }{6} \,.
\label{e13}
\end{eqnarray}

In equation~(\ref{e11}), $\s^2$ is the leading, \emph{linear},
contribution to the variance of the density field, $\s^2 \equiv \lan
\de_1^2 \ran$. The quantity $S_{3\de}$ denotes the skewness of the
density field,
\be
S_{3\de} = \f{\lan \de^3 \ran}{\lan \de^2 \ran^2} =  
\f{3 \lan \de_1^2 \de_2 \ran}{\s^4} + 
{\cal O}(\s^2) \,,
\label{e14}
\ee
and $S_{3\te}$ is the skewness of the velocity divergence field
defined in an analogous way. The quantities $\Sigma_2$ and $\Sigma_4$
are given by
\be 
\Sigma_{2} = \frac{ \lan \delta_{2} \theta_{2} \ran_c - \lan
\theta_{2}^{2} \ran_c + \lan \delta_{1} \delta_{3} \ran_c - \lan
\theta_{1} \theta_{3} \ran_c}{\sigma^{4}}
\label{e16}
\ee
and
\be
\Sigma_{4} = \frac{3 \lan \delta_{1}^{2} \delta_{2} \theta_{2} \ran_c
- 3 \lan \theta_{1}^{2} \theta_{2}^{2} \ran_c + \lan \delta_{1}^{3}
\delta_{3} \ran_c - \lan \theta_{1}^{3} \theta_{3}
\ran_c}{\sigma^{6}} .
\label{e17}
\ee
In the expressions above, the symbol $\lan \cdot \ran_c$ stands for
the connected (reduced) part of the moments.

In the derivation of the coefficients $a_i$, no assumption was made
about the particular forms of $\de_j$ and $\te_j$, except the linear
theory result $\de_1 = \te_1$.  Consequently, the inverse relation
for the mean of $\te$ given $\de$, can immediately be written by
exchanging the symbols $\de$ and $\te$ in equations~(\ref{e3})
and~(\ref{e11})--(\ref{e17}).  Thus if we express the function
$\Theta \equiv \lan \te \ran|_{\de}$ in the following form

\be
\Theta\left(\de\right) = r_1 \de + r_2 (\de^2 - \s_\de^2) + r_3 \de^3 
\,, \label{e19}
\ee
where $\s_\de^2 \equiv \lan \de^2 \ran$ is the variance of the density
field, then the coefficients $r_i$ can be written as
\begin{eqnarray}
r_1
&=& 1 + \left[ \Sigma_2' + \f{(S_{3\te} - S_{3\de})\, S_{3\de} }{3} -
\f{\Sigma_4'}{2} \right] \s^2 \,,
\label{e21} \\
r_2
&=& \f{S_{3\te} - S_{3\de}}{6} \,,
\label{e22} \\
r_3
&=& \f{\Sigma_4' - (S_{3\te} - S_{3\de})\, S_{3\de} }{6} \,,
\label{e23}
\end{eqnarray}
where
\be
\Sigma_{2}' = \frac{ \lan \te_{2} \de_{2} \ran_c - \lan
  \de_{2}^{2} \ran_c + \lan \te_{1} \te_{3} \ran_c -
  \lan \de_{1} \de_{3} \ran_c}{\sigma^{4}}
\label{e24}
\ee
and
\be
\Sigma_{4}' = \frac{3 \lan \te_{1}^{2} \te_{2} \de_{2} \ran_c
- 3 \lan \de_{1}^{2} \de_{2}^{2} \ran_c + \lan \te_{1}^{3}
\te_{3} \ran_c - \lan \de_{1}^{3} \de_{3}
\ran_c}{\sigma^{6}} .
\label{e25}
\ee

Equations~(\ref{e19})--(\ref{e25}) are, strictly speaking, valid only
up to third order in $\s_\de$ and for typical values of $\de$, i.e.,
away from rare peaks.  The spatial average of $\te$ is given by $\int
\lan \te \ran|_{\de} p(\de)\, d\de = \int \Theta\left(\de\right)
p(\de)\, d\de$. The mean of the first two terms in
equation~(\ref{e19}) is exactly zero. The mean of the third term is
$r_3 \lan \de^3 \ran = r_3 S_{3\de} \s_\de^4 \sim \calO(\s_\de^4)$
(see eq.~[\ref{e14}]). Hence, our third-order formula~(\ref{e19})
fulfills the requirement that the average of the velocity divergence
vanishes up to the terms cubic in $\s_\de$, as expected.

By comparing the coefficients $r_i$ with the expressions
(\ref{e11})-(\ref{e17}) for the coefficients $a_i$, we see that the
only new quantities to be computed are $\lan \de_{2}^{2} \ran_c$ and
$\lan \de_{1}^{2} \de_{2}^{2} \ran_c$, appearing in $\Sigma_{2}'$ and
$\Sigma_{4}'$ respectively. Therefore, one can express $r_i$ as
combinations of $a_i$ and some residuals $D_i$. We uniquely define the
residuals $D_i$ in the following way.  Let us approximate the function
$\Theta$ by the perturbative inversion of the function $\Delta$,
equation~(\ref{e3}). The approximations of the coefficients $r_i$
obtained in this manner will be denoted by $n_i$. The exact values of
the coefficients are then given by a sum of the approximate values
$n_i$ and the residuals $D_i$,

\be 
r_i = n_i + D_i \,.
\label{e26}
\ee
Straightforward calculation of the coefficients $n_i$ yields
\begin{eqnarray}
n_1
&=& 2 - a_1 - 2 a_2^2 \s^2 ,
\label{e27} \\
n_2
&=& - a_2 \,,
\label{e28} \\
n_3
&=& - a_3 + 2 a_2^2  \,.
\label{e29}
\end{eqnarray}

We will now transform the coefficients $r_i$ into forms similar to the
coefficients $n_i$. The case of $r_2$ is trivial:
\be
r_2 = \f{S_{3\te} - S_{3\de}}{6} = - \f{S_{3\de} - S_{3\te}}{6} = -
a_2 \,,
\label{e30}
\ee
so comparing with equation~(\ref{e28}) we see that $D_2 = 0$.  The
calculation of $r_1$ and $r_3$ is more lengthy but
straightforward. The transformation of the coefficient $r_3$ is
simpler when we rewrite $\Sigma_4'$ in a form consisting of 
explicitly asymmetric and symmetric parts,
\be 
\Sigma_4'
= \frac{S_{4\te} - S_{4\de}}{4} - \frac{ 3 \lan \te_{1}^{2}
(\de_{2} - \te_{2})^2 \ran_c}{2 \s^6} \,.
\label{e31}
\ee
Here, $S_{4\de}$ is the kurtosis of the density field, 
\be
S_{4 \delta} = \frac{6 \lan \delta_{1}^{2} \delta_{2}^{2} \ran_c
+ 4 \lan \delta_{1}^{3} \delta_{3} \ran_c}{\sigma^{6}} + 
\calO(\s^2) \,,
\label{e32}
\ee
and $S_{4\te}$ is the kurtosis of the velocity divergence field
defined in an analogous way. 
The final result is
\begin{eqnarray}
r_1
&=& 2 - a_1 - 2 a_2^2 \s^2 + \tilde D_1 \s^2 \,,
\label{e34} \\
r_2
&=& - a_2 \,,
\label{e35} \\
r_3
&=& - a_3 + 2 a_2^2 + D_3 \,,
\label{e36}
\end{eqnarray}
where
\be
\tilde D_1 = \f{ 3 \lan \te_{1}^{2} (\de_{2} - \te_{2})^2 \ran_c}{2 \s^6} 
- \f{ \lan (\de_{2} - \te_{2})^2 \ran_c}{\s^4} 
- \f{5 (S_{3\de} - S_{3\te})^2}{18}
\label{e37}
\ee
and
\be
D_3 = \f{(S_{3\de} - S_{3\te})^2}{9}
- \f{ \lan \te_{1}^{2} (\de_{2} - \te_{2})^2 \ran_c}{2 \s^6} \,.
\label{e38}
\ee
Note that we have introduced $\tilde D_1 = D_1/\s^2$ since this
rescaled parameter, like $D_3$, does not depend on $\s$. 

The expressions for $\Delta$ and $\Theta$ up to {\em second}
order in perturbation theory can be derived by neglecting terms of
order $\s^3$. The second order expression for the mean $\te$ given
$\de$ is therefore
\be
\lan \te \ran|_{\de} = \de - a_2 (\de^2 - \s_\de^2) 
\label{e39}
\ee
(cf.~Bernardeau~1992).  This expression is identical to that obtained
by direct inversion of the second order expression for $\lan \de
\ran|_{\te}$. Only when third order corrections are included, does the
scatter around the mean introduce a bias in the relation when derived by
direct inversion. It is interesting to note that although $D_1$ and
$D_3$ introduce ${\cal O}(\s^3)$ corrections, they themselves are
constructed exclusively from the first and the second order solutions,
$\de_1 =\te_1$, $\de_2$ and $\te_2$.

\section{Numerical calculations of $\bmath{\lowercase{r_i}}$}
\label{sec-num}
All of the numerical results presented in this section are performed
for fields smoothed with a Gaussian filtering window of width $R$. All
the terms appearing in the expressions for $r_i$ have exactly the same
mathematical structure as the terms appearing in the expressions for
$a_i$. Therefore, the same arguments that we used in Paper~I to argue
that the parameters $a_i$ are almost $\Omega$-independent apply as
well to the case of $r_i$. For example, $r_2 = - a_2$, and in Paper~I
we explicitly derived the $\Omega$-dependence of $a_2$ and showed it
to be extremely weak. Thus, the following calculations of $r_i$ are
performed for a flat Universe, but we expect $r_i$ to be robust to the
value of $\Omega$. 

The calculation of the terms $\langle \delta_{2}^{2}
\rangle/\sigma^{4}$ and $\langle \delta_{1}^{2} \delta_{2}^{2}\rangle/
\sigma^{6}$ which contribute, respectively, to $\Sigma_{2}'$ and
$\Sigma_{4}'$ can be found in \L okas et al. (1995, 1996). The rest of
the terms appearing in the expressions for $r_i$ were calculated in
Paper~I.

\subsection{Power law spectra}

First we present results for the scale-free spectra of the form
\begin{equation}  \label{n2}
    P(k) = C k^{n}, \ \ \ -3 \le n \le 1.
\end{equation}
The coefficients $r_{2}$ and $r_{3}$ are then independent of the
normalization of the spectrum and the smoothing scale. The results are
straightforward to obtain from equations (\ref{e22})--(\ref{e23}) and
they are given in Table~\ref{tab-rn} for various spectral indices.

\begin{table}
\caption{The coefficients $r_{1}$, $r_{2}$, $r_{3}$ and $n_{3}$ 
as functions of the spectral index $n$ for scale-free power spectra
and Gaussian smoothing}

\label{tab-rn}
\begin{tabular}{rcrcr}
\multicolumn{1}{c}{$n$} &
\multicolumn{1}{c}{\ \ $r_{1}$}  &
\multicolumn{1}{c}{\ \ $r_{2}$}  &
\multicolumn{1}{c}{\ \ $r_{3}$}  &
\multicolumn{1}{c}{\ \ $n_{3}$} \\  \hline
$-3.0$  &  $\approx 1+ 0.3 \s^2$ & $-0.190$ & $0.0826$ & $0.0826$ \\
$-2.5$  &  $1 + 0.202 \s^2$   &  $-0.192$ & 0.0822 & 0.0822 \\
$-2.0$  &  $1 + 0.077 \s^2$  &  $-0.196$ & 0.0818 & 0.0821 \\
$-1.5$  &  $1 - 0.296 \s^2$  &  $-0.203$ & 0.0812 & 0.0822 \\
$-1.0$  &  --   &  $-0.213$ & 0.0806 & 0.0835 \\
$-0.5$  &  --   &  $-0.227$ & 0.0797 & 0.0865 \\
   $0$  &  --   &  $-0.246$ & 0.0783 & 0.0928 \\
 $0.5$  &  --   &  $-0.270$ & 0.0756 & 0.1051 \\
 $1.0$  &  --   &  $-0.301$ & 0.0707 & 0.1283  
\end{tabular}
\end{table}

By comparison of $r_{2}$ and $r_{3}$ given by equations
(\ref{e35})--(\ref{e36}) with the approximate values $n_{2}$ and
$n_{3}$ given by equations (\ref{e28})--(\ref{e29}) we find that for
$n>-3$, the residual $D_{2}=0$ but $D_{3} \neq 0$.  Only for $n=-3$ we
find that $r_{3}$ exactly equals to $n_{3}$ and $D_{3}=0$. We present
$n_3$ in the last column of Table~\ref{tab-rn}. We see that the
approximate values $n_{3}$ diverge from the exact ones, $r_{3}$, more
significantly for higher spectral indices.

In order to obtain $r_{1}$ we use equation (\ref{e21}) and get for the 
unsmoothed fields
\begin{equation}  \label{n3}
    \Sigma_{2}' = -\frac{1681}{4410} - h(n) \approx -0.4
\end{equation}
where $h(n)$ is the part weakly dependent on $n$ which contributes 
roughly 10\%
to the value of $\Sigma_{2}'$. Using this result we estimate the value of 
the coefficient $r_{1}$ in this case to be
\begin{equation}  \label{n4}
    r_{1} \approx 1 + 0.3 \sigma^{2}.
\end{equation}
The value of the residual $\tilde{D}_{1} = D_{1}/\sigma^{2}$ given by 
equation (\ref{e37}) is for the unsmoothed fields independent of $n$ and 
equal to $-32/2205 = -0.0145$.

When smoothing is applied we are restricted to the spectral indices
$n<-1$ as in the case of $a_{1}$ (see Paper~I). The case of $n=-3$
with smoothing corresponds to the same case with no smoothing (because
for such a spectrum the dominant contribution comes from the small
wave-numbers at which the window function equals to unity), hence
$r_1$ is given by equation~(\ref{n4}). For $n=-2$ the calculations can
be performed analytically and we find
\begin{equation}  \label{n5}
     \Sigma_{2}' = -\frac{29}{196} \pi = -0.465
\end{equation}
hence
\begin{equation}  \label{n6}
    r_{1} = 1 + 0.0768 \sigma^{2}.
\end{equation}

For half-integer values of spectral indices we calculated the values
of $r_{1}$ numerically. The results are given in the second column of
Table~\ref{tab-rn}. Comparing equation (\ref{e34}) with equation
(\ref{e27}) we obtain the values of $\tilde{D}_{1} =
D_{1}/\sigma^{2}$. We find that in this range of spectral indices the
coefficient $n_{1}$ is a good approximation of $r_{1}$ that is the
correction introduced by $\tilde{D}_{1}$ is very small. Therefore, in
Table~\ref{tab-rn} we do not present the values of $n_1$.

\subsection{The standard CDM spectrum}

As an example of a scale-dependent power spectrum we consider the standard 
CDM (with coefficients given by Efstathiou, Bond \& White 1992) normalized 
so that the linear rms fluctuation in spheres of radius $R = 8 h^{-1}$ Mpc 
is equal to unity. Given $a_{2}$ the coefficient $r_{2}$ is obtained 
immediately from equation (\ref{e35}). 

In calculating $r_{3}$ we use the effective index
\begin{equation}  \label{n7}
    n_{\rm eff} = - \frac{R}{\sigma^{2}} \frac{{\rm d} \sigma^{2}(R)}
    {{\rm d} R} - 3
\end{equation}
which measures the slope of the power spectrum at smoothing scale $R$.
Then the values of $r_{3}$ at a given scale are found by interpolating
the values known for scale-free spectra (Table~\ref{tab-rn}) at the
effective index corresponding to that scale.

The values of $\Sigma_{2}'$ that contribute to $r_{1}$ cannot however
be calculated using the effective index method. We therefore calculate
the correction to unity in $r_{1}$ by numerical integration of the
relevant formulae (see Paper I). As in the case of $a_{1}$ we benefit
from the fact that the slope of the CDM spectrum approaches $-3$ at
large wave-numbers and thus the weakly nonlinear correction to the
value of $r_{1}$ converges even for scales where the effective
spectral index is close to $+1$.

\begin{figure}
\begin{center}
    \leavevmode
    \epsfxsize=7cm
    \epsfbox[0 200 550 710]{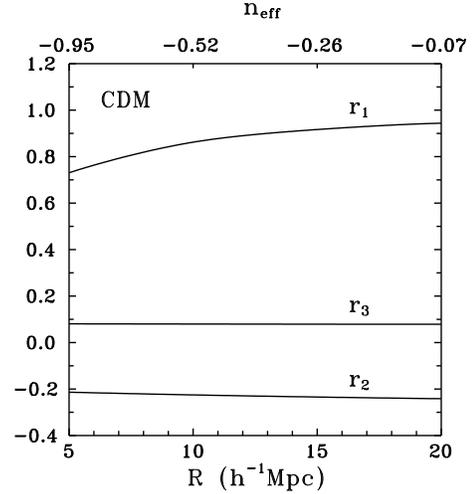}
\end{center}
\caption{The coefficients $r_{1}$, $r_{2}$ and $r_{3}$ for the standard CDM
spectrum normalized to (top hat) $\sigma_{8} = 1$ in the weakly
nonlinear range of Gaussian smoothing scales. The values of the
effective index corresponding to each of the smoothing scales are also
plotted.}

\label{fig-CDM}
\end{figure}

The results for the CDM spectrum in the weakly nonlinear range of
scales are presented in Figure~\ref{fig-CDM}. In the values of $r_{1}$
we incorporated the linear variances $\sigma^{2}$ determined by the
normalization to $\sigma_{8} = 1$. The Figure also shows the values of
the effective index corresponding to each of the smoothing scales.

\section{Scatter in the $\bmath{\de}$--$\bmath{\te}$ relation}
\label{sec-scatter}
In Paper~I we have computed mean $\de$ given $\te$, $\lan \de
\ran|_{\te}$. Analogous calculations lead to a formula for the
conditional variance, $\s^2|_{\te} \equiv \lan (\de - \lan \de
\ran|_{\te})^2 \ran|_{\te}$; we present them in brief in
Appendix~A. The result is

\be
\s^2|_{\te} = b_0 \s_\te^4 + b_2 \s_\te^2 \te^2 + \calO(\s_\te^5)\,,
\label{s1}
\ee
where

\be
b_0 = \f{\lan (\de_2 - \te_2)^2 \ran}{\sigma^4} - 
\f{\lan \te_1^2 (\de_2 - \te_2)^2 \ran}{2 \sigma^6} + 
\f{(\Delta S_3)^2}{18}
\label{s2}
\ee
and

\be
b_2 =  \f{\lan \te_1^2 (\de_2 - \te_2)^2 \ran}{2 \sigma^6} -
\f{(\Delta S_3)^2}{9} \,.
\label{s3}
\ee
Here, $\Delta S_3 \equiv S_{3\de} - S_{3\te}$.

In the present paper we have computed mean $\te$ given $\de$, $\lan
\te \ran|_{\de}$. The conditional variance in this case, $\s^2|_{\de}
\equiv \lan (\te - \lan \te \ran|_{\de})^2 \ran|_{\de}$, may be
similarly obtained from equations~(\ref{s1})--(\ref{s3}) by exchanging
the symbols $\de$ with $\te$. Unlike the coefficients of the mean
trend, equations~(\ref{e11})--(\ref{e17}), the coefficients $b_0$ and
$b_2$ are invariants with respect to this exchange.  Therefore,

\be
\s^2|_{\de} = b_0 \s_\de^4 + b_2 \s_\de^2 \de ^2 + \calO(\s_\de^5)\,,
\label{s4}
\ee
with $b_0$ and $b_2$ given by equations~(\ref{s2}) and~(\ref{s3})
respectively. 

The coefficients $b_0$ and $b_2$ may be written in terms of the
residuals $\tilde D_1$ and $D_3$, equations~(\ref{e37})
and~(\ref{e38}):

\be
b_0 = - \tilde D_1 - 2 D_3 
\label{s5}
\ee
and
\be
b_2 = -D_3 \,.
\label{s6}
\ee

Thus the presence of the residuals in
equations~(\ref{e34})--(\ref{e36}) for the coefficients $r_i$ is a
direct consequence of the scatter in the relation between $\de$ and
$\te$. Were the relation entirely deterministic, it would be
describable by one function, so then $\Theta = \Delta^{-1}$,
$r_i = n_i$ and there would be no residuals. However, since in weakly
nonlinear regime the values of pairs ($\de$,$\te$) form an elongated
set of some scatter, it is not surprising that averaging along
different coordinates gives different curves.

As typically $\de \sim \te \sim \s$, we have $\s^2 \de^2 \sim \s^2
\te^2 \sim \s^4$ and both terms entering the formulas~(\ref{s1})
or~(\ref{s4}) for the conditional variances are of the order of
$\s^4$. The rms value of the scatter in the $\de$--$\te$ relation --
the square root of the conditional variance -- is therefore of the
order of $\s^2$. A good measure of the elongation of a two-dimensional
set of points ($\de$,$\te$) is the ratio of its scatter to a typical
value of $\de$ or $\te$. In our case this ratio is $\sim \s$. When $\s
\to 0$, the ratio thus tends to zero, as expected: in the limit of
linear theory the relation between $\de$ and $\te$ is deterministic.

We present the values of $b_0$ and $b_2$ as functions of the spectral
index $n$ in Table~\ref{tab-scatter}. We see that $b_0$ and $b_2$ are
always positive and the conditional variance is positive-definite, as
required.

\begin{table}
\caption{The coefficients $b_{0}$ and $b_{2}$ as functions of the 
spectral index $n$ for scale-free power spectra and Gaussian
smoothing}
\label{tab-scatter}
\begin{tabular}{rrc} 
\multicolumn{1}{c}{$n$} & 
\multicolumn{1}{r}{$b_0$\ \ \ \ } & 
\multicolumn{1}{c}{$b_2$}  \\  \hline
-3.0  &  $\f{32}{2205} \simeq 0.0145$  &  0        \\
-2.5  &  0.0155  &  $3.61 \cdot 10^{-5}$ \\
-2.0  &  0.0193  &  $2.66 \cdot 10^{-4}$ \\
-1.5  &  0.0288  &  $1.02  \cdot 10^{-3}$ \\
-1.0  &  0.0517  &  $2.86 \cdot 10^{-3}$  \\
-0.5  &  0.116\ \ &  $6.77 \cdot 10^{-3}$  \\
 0.0  &  0.378\ \ &  $1.45 \cdot 10^{-2}$   \\
 0.5  &  -- \ \ \ &  $2.95 \cdot 10^{-2}$    \\
 1.0  &  -- \ \ \ &  $5.76 \cdot 10^{-2}$
\end{tabular}
\end{table}

It is rather astonishing that the formulas for the coefficients $b_0$
and $b_2$ for the conditional variance are constructed exclusively
from second-order perturbative contributions for $\de$ and $\te$. In
section~2 we have shown that up to second order $\Theta =
\Delta^{-1}$; the residuals $D_i$ appear only when third-order
corrections to the mean trend are included. Still, $D_i$'s and $b_i$'s
contain only second-order terms. Why? The explanation of this apparent
paradox is provided by Chodorowski (1997). In this work, for
unsmoothed fields, the weakly nonlinear density has been shown to be a
local function of the two velocity scalars: the expansion (velocity
divergence) and the shear,

\be 
\de(\bfx) =  \te(\bfx) + \f{4}{21} \left[\te^2(\bfx) - 
\f{3}{2} \Sig^2(\bfx) \right] + \calO(\s^3)
\label{s7} 
\ee
(cf.~Mancinelli \& Yahil 1995 and Catelan \etal 1995). Here, the
shear scalar $\Sig$ is

\be
\Sig \equiv 
\left( \Sig_{ij} \Sig_{ij} \right)^{1/2} \,,
\label{s8} 
\ee
where

\be
\Sig_{ij} \equiv \f{1}{2} \left( v_{i,j} + v_{j,i} \right) - 
\f{1}{3} \te \de_{ij}
\label{s9} 
\ee
and $v_{i,j}$ are velocity derivatives. Thus, the scatter in the
$\de$--$\te$ relation comes from the shear term in
equation~(\ref{s7}). The mean trend is

\be
\lan \de \ran|_{\te} = \te + 
\f{4}{21} \left[ \te^2 - \f{3}{2} \lan \Sig^2 \ran|_{\te} \right]
+ \calO(\s^3) \,.
\label{s10} 
\ee
It reproduces up to second order equation~(\ref{e3}) since
at the lowest order $\te$ and $\Sig$ are statistically independent,
hence $\lan \Sig^2 \ran|_{\te} = \lan \Sig^2 \ran = (2/3) \s^2$ (see
Chodorowski 1997 for details). The conditional variance is
\begin{eqnarray}
\s^2|_{\te} 
&=& 
\f{4}{49} \left.\left\lan \left( \Sig^2 - \lan \Sig^2 \ran|_{\te} + 
\calO(\s^3) \right)^2 \right\ran\right|_{\te} \nonumber \\
&=&
\f{4}{49} \left\lan \left( \Sig^2 - \lan \Sig^2 \ran \right)^2 
\right\ran + \calO(\s^5) \,.
\label{s11} 
\end{eqnarray} 

The linear term in equations~(\ref{s7}) and~(\ref{s10}), $\te$, has
cancelled out in the above formula for the conditional variance.
Therefore, third-order contributions to equation~(\ref{s7}) result in
contributions to the conditional variance that are already of the
order $\s^5$. This is in contrast to the formula for weakly nonlinear
growth of ordinary variance,

\be
\lan \de^2 \ran - \lan \de_1^2 \ran = \lan \de_2^2 \ran + 
2 \lan \de_1 \de_3 \ran \,,
\label{s12} 
\ee 
which does involve a third-order contribution due to the presence of
the linear term. 

The above consideration is strictly valid only for unsmoothed fields,
but the linear terms in the conditional variance will also cancel out
for smoothed fields and the lowest order contribution to it will be
constructed exclusively from second-order terms.

The ensemble average on the RHS of equation~(\ref{s11}) can be
performed: the result is 

\be
\s^2|_{\te}  = \f{32}{2205} \s^4 + \calO(\s^5) \,.
\label{s16} 
\ee
The case of unsmoothed fields corresponds to the case of smoothed
fields with the spectral index $n = -3$. From Table~\ref{tab-scatter} we
see that for $n = -3$ the parameters $b_0 = 32/2205$, $b_2 = 0$ and
equation~(\ref{s1}) coincides with the above formula. Thus,
equation~(\ref{s1}) is an extension of the result of Chodorowski
(1997) for the case of smoothed fields.

In this work we are mostly interested in the conditional variance
$\lan (\te - \lan \te \ran|_{\de})^2 \ran|_{\de}$. In
Table~\ref{tab-scatter} we see that $b_2 \ll b_0$ always. For mildly
nonlinear fields ($\s_\de \la 1$) this implies that in
equation~(\ref{s4}) for $\s^2|_{\de}$ the term quadratic in $\de$ is
negligible with respect to the constant one. [The coefficient $b_2$ is
\emph{exactly} zero in the case of a top-hat smoothing (Chodorowski et
al. 1998)]. Furthermore, the actual $\de$-dependence of the
conditional variance must originate from the terms that are of order
higher than $\s_\de^4$. In Appendix~A we show that the next-order term
is $\propto \s_\de^4 \de$ which is of the order of $\s_\de^5$; other
terms are already $\calO(\s_\de^6)$. This is indeed observed in N-body
simulations, where the estimated variance can be well described by the
formula (Chodorowski et al. 1998)

\be
\s^2|_{\de}^{(NB)} = b_0^{(NB)} \s_\de^4 + b_1^{(NB)} \s_\de^4 \de \,,
\label{s17}
\ee

with $b_1^{(NB)} \simeq 0.8 b_0^{(NB)}$. In principle, the calculation
of the coefficient $b_1$ involves fourth-order perturbation theory. If
density is a local function of the velocity divergence and the shear
up to \emph{third} order, similar arguments to these presented after
equation~(\ref{s7}) may be used to argue that the formula for $b_1$
should be composed from at most third-order perturbative
contributions. Even if this is true, however, we find that the moments
$\s^{-6} \lan \de_2 (\de_2 - \te_2)^2) \ran$ and $\s^{-6} \lan \te_1
(\de_2 - \te_2)(\de_3 - \te_3) \ran$ will enter into this formula, of
extra complexity in comparison to any moments computed here. Given all
that, in this paper we will not try to predict the value of $b_1$.

\section{Summary}
\label{sec-summ}
We have derived a local relation between the divergence of the weakly
non-linear peculiar velocity field and the corresponding density
contrast field. Specifically, we have computed the mean value of the
velocity divergence given density up to third order in perturbation
theory, assuming Gaussian initial conditions. Our perturbative
calculation yields a third-order `expansion' of the mean value of the
divergence in terms of the density contrast. The coefficients of this
expansion have been explicitly calculated for scale-free and standard
CDM power spectra, for Gaussian smoothing of the fields in question.
In the case of CDM, the coefficients depend weakly on the smoothing
scale. Moreover, they are expected to depend very weakly on $\Omega$.
It is interesting that the value of the linear coefficient in the
relation differs from unity -- the value predicted in linear
theory. The corrective term to this value is proportional to the
variance of the density field. The same form of the corrective term is
found in the relation between the mean velocity divergence and the
density resulting from the Zel'dovich approximation
(Chodorowski~1998).

In order to assess the tightness of the relation, we have computed the
scatter in the velocity divergence around its mean value given
density. We find that, at least up to third order in perturbation
theory, the scatter is almost independent of the density contrast. The
rms value of the scatter relative to the rms value, $\s_\te$, of the
divergence is approximately $b_0^{1/2} \s_\te$ where
$b_0=0.015$--$0.37$ for scale free power spectra with $n$ ranging from
$-3$ to $0$. Over scales larger than about $5 \hmpc$, this scatter is
small relative to that introduced by observational errors in analyses
of redshift surveys and catalogs of peculiar velocities, such as the
density--density or velocity--velocity comparisons. The main source of
these errors are the sparse and non-uniform sampling of the velocity
data and the scatter in the distance indicators (e.g. the Tully-Fisher
relation). As an example, let us consider the {\it IRAS}--{\sc potent}
comparison (Sigad \etal 1998), employing Gaussian smoothing length of
$12 \hmpc$. At this scale, $\s_\te$ is approximately $0.3$, hence the
ratio of the rms value of the scatter around our relation to the rms
value of the divergence lies in the range $0.04$--$0.18$. In contrast,
the rms value of the observational scatter is approximately $0.21$
(estimated from the mock catalogs in the `standard volume'), thus the
observed ratio is as high as $0.7$. N-body simulations (Nusser \etal
1991, Mancinelli \etal 1994, Chodorowski \etal 1998; cf.~also Sigad
\etal 1998) also demonstrate that the scatter around relations of this
type is relatively small. The relation and the formalism presented
here can be valuable for quantifying the redshift space anisotropies
in the non-linear regime. This is particularly important for probing
small scales in the planned SDSS and 2dF surveys. We intend to use the
relation for estimating the amount of anisotropy of structure in
redshift space.

\section*{Acknowledgments}
This research has been supported in part by the Polish State Committee
for Scientific Research grants No.~2.P03D.008.13 and 2.P03D.004.13.

\onecolumn

\appendix
\section{Calculation of the conditional variance}
Here we describe the calculation of the variance $\s^2|_{\te} = \lan
\de^2 \ran|_{\te} - \lan \de \ran|_{\te}^2$. The derivation is
analogous to the derivation of the mean trend, presented in Paper~I,
therefore we will sketch it in brief.  If $p(\de,\te)$ is the joint
probability distribution function (PDF) for $\de$ and $\te$, then
mean $\de^{2}$ given $\te$ is 
\be 
\lan \de^{2} \ran |_{\te} = \f{\int \de^{2} p(\de,\te)\, 
{\rm d}\de}{p(\te)} \,.
\label{a4}
\ee

The quantity $\calN \equiv \int \de^{2} p(\de,\te)\, {\rm d}\de$ can be
expressed as 
\be \calN = \f{1}{2\pi} \int {\rm e}^{-{\rm i}s\te}
 \left. \f{\p^{2}}{\p({\rm i}t)^{2}} \Phi({\rm i}t,{\rm i}s) 
\right|_{t=0} \, {\rm d}s\,,
\label{a6} 
\ee 
where $\Phi$ is the characteristic function of the joint PDF. It is
related to the cumulant generating function, $\calK$, by the equation

\be 
\Phi({\rm i}t,{\rm i}s) = \exp{[\calK({\rm i}t,{\rm i}s)]}
\,. \label{a7} 
\ee 

The cumulants, $\kappa_{mn}$, from which $\calK$ is constructed, 

\be
\calK = \sum_{(m,n) \ne (0,0)}^\infty \f{\kappa_{mn}}{m! n!}  ({\rm i}t)^m
({\rm i}s)^n \,, \label{a8} 
\ee 
are given by the {\em connected} part of the joint moments

\be \kappa_{mn} = \lan
\de^m \te^n \ran_{c} \,.
\label{a9}
\ee
Using equations~(\ref{a7}) and (\ref{a8}) we obtain
\be
\left. \f{\p^{2}}{\p({\rm i}t)^{2}} \Phi({\rm i}t,{\rm i}s) \right|_{t=0} =
\left[ 
\sum_{n=0}^\infty \f{\kappa_{2n}}{n!} ({\rm i}s)^n + 
\sum_{n,p=0}^\infty \f{\kappa_{1n} \kappa_{1p}}{n! p!} ({\rm i}s)^{n+p} 
\right] \,
\exp\!\left[\sum_{n=1}^\infty \f{\kappa_{0n}}{n!} ({\rm i}s)^n \right] \,.
\label{a10}
\ee
By defining $z = \kappa_{02}^{1/2} s$, $\mu = {\de}/{\s_\de} =
 {\de}/{\kappa_{20}^{1/2}}$, $\nu = {\te}/{\s_\te} =
 {\te}/{\kappa_{02}^{1/2}}$, and the standard cumulants
$\lambda_{mn} = \kappa_{mn} / (\kappa_{20}^{m/2} \kappa_{02}^{n/2})$, 
we find that
\be
\calN = \f{1}{2\pi} \f{\kappa_{20}}{\kappa_{02}^{1/2}}
\int_{-\infty}^\infty {\rm d}z\, e^{-\f{1}{2}(z^2 + 2{\rm i}\nu z)}
\left[
\sum_{n=0}^\infty \f{\lambda_{2n}}{n!} ({\rm i}z)^n + 
\sum_{n,p=1}^\infty \f{\lam_{1n} \lam_{1p}}{n! p!} ({\rm i}z)^{n+p}
\right] 
\exp\!\left[\sum_{n=3}^\infty \f{\lam_{0n}}{n!} ({\rm i}z)^n \right] .
\label{a14}
\ee

In weakly nonlinear regime the standard cumulants obey the following
scaling hierarchy (Fry 1984, Bernardeau 1992)

\be
\lam_{mn} = S_{mn} \s^{m+n-2} + {\cal O}(\s^{m+n}) 
\label{a15}
\ee
where $\s$ is the linear variance of $\de$ or, equivalently, of $\te$
(recall that at linear order $\de = \te$). The series in
equation~(\ref{a14}) are thus power series in a small parameter $\s$
and we can truncate them at some order $p$ neglecting contributions
which are  of order $> \s^{p}$. The leading-order formula for the
conditional variance is obtained by keeping the terms up to the order
of $\s^2$. Integrating the resulting expression yields
\begin{eqnarray}
\calN
&=& \f{1}{\sqrt{2\pi}} 
\f{\kappa_{20}}{\kappa_{02}^{1/2}} e^{-\f{1}{2} \nu^2} \times
\nonumber \\
&\ & \left[H_0(\nu) + \lam_{11}^2 H_2(\nu) + \lam_{21} H_1(\nu) + 
\left(\lam_{12} + \f{\lam_{03}}{6}\right) H_3(\nu) + 
\f{\lam_{03}}{6} H_5(\nu) +
\right.
\nonumber \\
&\ &
\f{\lam_{22}}{2} H_2(\nu) +
\left(
\f{\lam_{13}}{3} + \f{\lam_{04}}{24} + 
\f{\lam_{21}\lam_{03}}{6} + \f{\lam_{12}^2}{4} 
\right) H_4(\nu) +  
\nonumber \\
&\ &
\left. \left( \f{\lam_{04}}{24} + \f{\lam_{12}\lam_{03}}{6} +  
\f{\lam_{03}^2}{72} \right) H_6(\nu) + 
\f{\lam_{03}^2}{72} H_8(\nu)\right] \,,
\label{a18}
\end{eqnarray}
where $H_{n}$ are the $n$-th order Hermite polynomials.

We now turn to calculating the PDF $p(\theta)$ in equation~(\ref{a4}).
Computing it in a similar way as $\calN$ we \emph{rederive} the
so-called Edgeworth expansion for the variable $\nu$
(Longuet-Higgins~1963, Bernardeau \& Kofman 1995, Juszkiewicz \etal
1995),

 \be p(\nu) = \frac{1}{\sqrt{2 \pi}} \ {\rm
e}^{-\nu^{2}/2} \left[ 1 + \frac{1}{6} \lam_{03} H_{3}(\nu) \right.  +
\frac{1}{24} \lam_{04} H_{4}(\nu) \left. + \frac{1}{72} \lam_{03}^2
H_{6}(\nu) \right] \,.
\label{a19}
\ee
{}From equations~(\ref{a18}) and~(\ref{a19}), after some algebra we
obtain

\be
\lan \mu^2 \ran|_{\nu} = \calS_0(\nu) + \calS_1(\nu) + \calS_2(\nu)\,,
\label{a20}
\ee
where

\be 
\calS_0(\nu) = 1 - \lam_{11}^2 + \lam_{11}^2 \nu^2 \,,
\label{a25}
\ee
\be
\calS_1(\nu) = (\lam_{21} - 3\lam_{12} + 2\lam_{03}) \nu + 
(\lam_{12} - \lam_{03}) \nu^3 \,,
\label{a26}
\ee
and
\begin{eqnarray}
\calS_2(\nu) 
&=&  
- \f{ \lam_{22} - 2\lam_{13} + \lam_{04} }{2} + 
\f{ 2\lam_{21} \lam_{03} + 3 \lam_{12}^2 - 
10 \lam_{12} \lam_{03} + 5 \lam_{03}^2 }{4} + 
\nonumber \\
&\ &
\left(  
\f{ \lam_{22} - 4 \lam_{13} + 3 \lam_{04} }{2} +
\f{ -\lam_{21} \lam_{03} - 3 \lam_{12}^2 + 
12 \lam_{12} \lam_{03} - 8 \lam_{03}^2 }{2}   
\right) \nu^2 +
\nonumber \\
&\ &
\left( 
\f{ \lam_{13}  - \lam_{04} }{3} + 
\f{ \lam_{12}^2 - 6 \lam_{12} \lam_{03} + 5 \lam_{03}^2 }{4} 
\right) \nu^4
\,.
\label{a27}
\end{eqnarray}
Note that $\calS_k(\nu)$ are of the order of $\s_{}^k$.
To obtain $\lan \mu^2 \ran|_{\nu} - \lan \mu \ran|_{\nu}^2$ we use the
expression for $\lan \mu \ran|_{\nu}$  given explicitly in
Paper~I. The result is
\begin{eqnarray}
\lan \mu^2 \ran|_{\nu} - \lan \mu \ran|_{\nu}^2 
&=&  
1 - \lam_{11}^2 - \f{\lam_{22} -2\lam_{13} + \lam_{04}}{2} + 
\f{ \lam_{21} \lam_{03} + \lam_{12}^2 - 
4 \lam_{12} \lam_{03} + 2\lam_{03}^2 }{2} + 
\nonumber \\
&\ &
( \lam_{21} - 2 \lam_{12} + \lam_{03} ) \nu +
\nonumber \\
&\ &
\left( 
\f{ \lam_{22} - 2\lam_{13} + \lam_{04} }{2} 
- \f{ \lam_{21} \lam_{03} + 2\lam_{12}^2 
- 6\lam_{12} \lam_{03} + 3\lam_{03}^2 }{2}  
\right) \nu^2 \,. 
\label{a31}   
\end{eqnarray} 
Were the variables identical (to all orders), $\mu = \nu$, then the
cumulants would be $\lam_{11} = 1$, $\lam_{mn} = \lam_{(m+n)0}$, and
the conditional variance would be zero, as expected.  Returning to the
`physical' variables $\de$ and $\te$, using the leading-order
expressions for the standard cumulants, like

\be
\lam_{11}^2 = 1 - \f{\lan (\de_2 - \te_2)^2 \ran}{\s_{}^2} +
\calO(\s_{}^4) \,, 
\label{a34}
\ee
and recalling that $\s_\te = \s + \calO(\s_{}^3)$ we finally obtain

\be
\lan \de^2 \ran|_{\te} - \lan \de \ran|_{\te}^{\,2} = 
b_0 \s_{\te}^4 + b_2 \s_{\te}^2 \te ^2 + \calO(\s_{\te}^5) \,,
\label{a38}
\ee
where

\be
b_0 = \f{\lan (\de_2 - \te_2)^2 \ran}{\s_{}^4} - 
\f{\lan \te_1^2 (\de_2 - \te_2)^2) \ran}{2 \s_{}^6} + 
\f{(\Delta S_3)^2}{18}
\label{a39}
\ee
and

\be
b_2 =  \f{\lan \te_1^2 (\de_2 - \te_2)^2) \ran}{2 \s_{}^6} -
\f{(\Delta S_3)^2}{9} \,.
\label{a40}
\ee
In equation~(\ref{a31}) higher-order corrections to the values of
cumulants yield corrections to formula~(\ref{a38}) which are already
of the order of $\s_{\te}^6$. The only exception is the term linear in
$\nu = \te/\s_{\te}$ which yields a contribution of the order of
$\s_{\te}^5$. Therefore, for mildly nonlinear fields, the first departure
from the leading-order formula~(\ref{a38}) for the conditional variance
$\s^2|_{\te}$ should have the form $\propto \s_{\te}^4
\te$. Analogously, the expression for the conditional variance in the
reverse case, $\s^2|_{\de} = \lan \te^2 \ran|_{\de} - \lan \te
\ran|_{\de}^2$, should contain a higher-order term of the form $\propto
\s_{\de}^4 \de$.

\bsp
\end{document}